%% file: mybibfile.tex
\documentclass[10pt,twocolumn]{article} 
\usepackage{simpleConference}
\usepackage{times}
\usepackage{graphicx}
\usepackage{amssymb}
\usepackage[hyphens]{url}
\usepackage{hyperref}
\usepackage[numbers]{natbib}
\usepackage{subcaption}
\usepackage{amsthm}
\usepackage{multirow}
\usepackage{booktabs}
\usepackage{amsmath}
\usepackage[]{algorithm2e}
\usepackage{color}
\usepackage{multicol}
\usepackage{bm}
\usepackage{authblk}
\usepackage{commath}    
\usepackage[capitalize]{cleveref}
\usepackage[outline]{contour}

\crefname{section}{Sec.}{Secs.}
\Crefname{section}{Section}{Sections}
\Crefname{table}{Table}{Tables}
\crefname{table}{Tab.}{Tabs.}
\crefname{appendix}{Appx.}{Appxs.}
\crefname{figure}{Fig.}{Figs.}
\crefname{theorem}{Theorem}{Theorem}

\newtheoremstyle{break}
  {\topsep}{\topsep}%
  {\itshape}{}%
  {\bfseries}{}%
\theoremstyle{break}

\newcommand*\samethanks[1][\value{footnote}]{\footnotemark[#1]}

\newcommand{\mrshtwo}{\texttt{mrsh-v2}}
\newcommand{\mrshcf}{\texttt{mrsh-cf}}
\newcommand{\tlsh}{\texttt{TLSH}}
\newcommand{\ssdeep}{\texttt{ssdeep}}
\newcommand{\sdhash}{\texttt{sdhash}}

\title{Combining AI and AM –- Improving Approximate Matching through Transformer Networks}

\author[1,6]{Frieder Uhlig\thanks{Equal contribution}}
\author[1]{Lukas Struppek\samethanks}
\author[1]{Dominik Hintersdorf\samethanks}
\author[2]{Thomas Göbel}
\author[2]{Harald Baier}
\author[1,3,4,5,6]{Kristian Kersting}
\affil[1]{Department of Computer Science, Technical University of Darmstadt, Germany}
\affil[2]{University of the Bundeswehr Munich, Germany}
\affil[3]{Centre for Cognitive Science, Technical University of Darmstadt}
\affil[4]{German Research Center for Artificial Intelligence (DFKI), Germany}
\affil[5]{Hessian Center for AI (hessian.AI), Germany}
\affil[6]{SEC Consult Unternehmensberatung GmbH, Germany}

\makeatletter
\newcommand{\printfnsymbol}[1]{%
  \textsuperscript{\@fnsymbol{#1}}%
}

\begin{document}
\maketitle
\thispagestyle{empty}
\let\thefootnote\relax\footnotetext{Published at DFRWS USA 2023 as a conference paper.}

\input{sections_arxiv/0_abstract}

\input{sections_arxiv/1_introduction}

\input{sections_arxiv/2_background}
\input{sections_arxiv/3_experiments}
\input{sections_arxiv/4_discussion}

\input{sections_arxiv/5_conclusion}

\bibliographystyle{plainnat}
\bibliography{mybibfile}

\cleardoublepage
\newpage

\end{document}

%% file: sections_arxiv/0_abstract.tex
\begin{abstract}
Approximate matching is a well-known concept in digital forensics to determine the similarity between digital artifacts.  An important use case of approximate matching is the reliable and efficient detection of case-relevant data structures on a blacklist (e.g., malware or corporate secrets), if only fragments of the original are available. For instance, if only a cluster of indexed malware is still present during the digital forensic investigation, the approximate matching algorithm shall be able to assign the fragment to the blacklisted malware.  However, traditional approximate matching functions like \tlsh\ and \ssdeep\ fail to detect files based on their fragments if the presented piece is relatively small compared to the overall file size (e.g., like one-third of the total file). A second well-known issue with traditional approximate matching algorithms is the lack of scaling due to the ever-increasing lookup databases.  In this paper, we propose an improved matching algorithm based on transformer-based models from the field of natural language processing.  We call our approach \textit{Deep Learning Approximate Matching (DLAM)}. As a concept from artificial intelligence, DLAM gets knowledge of characteristic blacklisted patterns during its training phase.  Then DLAM is able to detect the patterns in a typically much larger file, that is DLAM focuses on the use case of fragment detection.  Our evaluation is inspired by two widespread blacklist use cases:  the detection of malware (e.g., in JavaScript) and corporate secrets (e.g., pdf or office documents).  We reveal that DLAM has three key advantages compared to the prominent conventional approaches \tlsh\ and \ssdeep. First, it makes the tedious extraction of known to be bad parts obsolete, which is necessary until now before any search for them with approximate matching algorithms. This allows efficient classification of files on a much larger scale, which is important due to exponentially increasing data to be investigated.  Second, depending on the use case, DLAM achieves a similar (in case of \mrshcf\ and \mrshtwo) or even significantly higher accuracy (in case of \ssdeep\ and \tlsh) in recovering fragments of blacklisted files. For instance, in the case of JavaScript files, our assessment shows that DLAM provides an accuracy of 93\% on our test corpus, while \tlsh\ and \ssdeep\ show a classification accuracy of only 50\%. Third, we show that DLAM enables the detection of file correlations in the output of \tlsh\ and \ssdeep\ even for fragment sizes, where the respective matching function of \tlsh\ and \ssdeep\ fails.
\end{abstract}

%% file: sections_arxiv/1_introduction.tex
\section{Introduction}\label{sec:introduction}
Digital forensics comprises the analysis and interpretation of digital artifacts. One of the biggest challenges facing digital forensic investigation is coping with the vast number of files to be processed. Hashing algorithms have become an indispensable part of computer science, since the computed hashes can be used as unique identifiers to compare digital artifacts. For traditional cryptographic hashes, such as MD5 and SHA-2, even changing a single bit already alters the hash value significantly due to their diffusion property (intended for one-way functions), which obscures the relationship between the input and its corresponding hash value. Cryptographic hashes are therefore unsuitable for recognizing similar artifacts. Fuzzy hashing, on the other hand, breaks the cryptographic diffusion and thus establishes a relationship between the original file and its corresponding hash. Through approximate matching, the similarity of two fuzzy hashes can be determined.

The concept of approximate matching is not limited to traditional fuzzy hashes. In the image domain, similar approaches based on neural networks to extract meaningful features from images are used for image retrieval \citep{liu2016_image_retrieval, Zhao_2015_CVPR} or detecting illegal content \citep{apple_tech_summary, struppek21_neural_hash}.

\begin{figure*}[ht]
     \centering
     \begin{subfigure}[b]{0.49\textwidth}
         \centering
         \includegraphics[height=5.5cm]{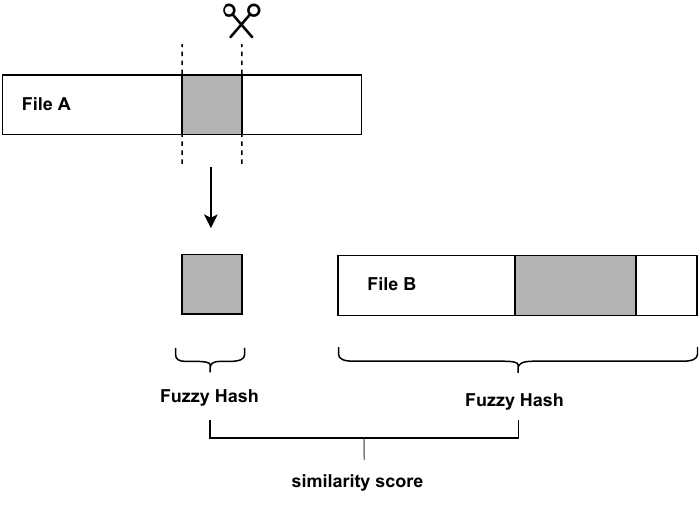}
         \caption{Fragment detection through conventional approximate matching.}
         \label{fig:anomaly_detection_conventional}
     \end{subfigure}
     \hfill
     \begin{subfigure}[b]{0.49\textwidth}
         \centering
         \includegraphics[height=5.5cm]{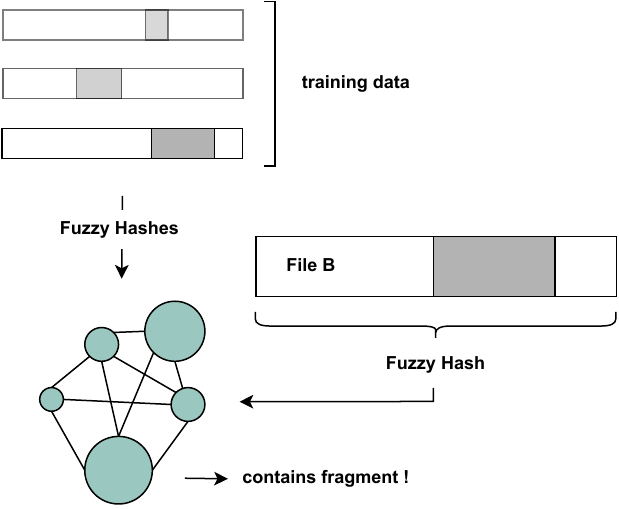}
          \caption{Fragment detection through DLAM.}
         \label{fig:anomaly_detection_dlam}
     \end{subfigure}
        \caption{Fragment detection through conventional approximate matching and through DLAM. In conventional approximate matching, the exact fragment that needs to be detected has to be extracted first in a costly operation. Through DLAM, models are trained to perceive the fragment as a common denominator and classify files based on its presence.}
        \label{fig:DLAM}
\end{figure*}

A recent survey by \citet{singh_2021} shows that 65\% of digital forensic investigators are actively using fuzzy hashes during their investigation processes. Fuzzy hashing is primarily used to identify related documents, but they are also used in other contexts which are based on similarity search. Examples are data loss prevention \citep{goebel_2021}, IoT device identification \citep{charyyev_2020, goebel_2022}, genome sequencing \citep{healy_2011} or biometric template protection \citep{Ong2014ImprovedFH}. However, one main application for fuzzy hashes remains malware detection and analysis~\citep{diaz_why_2020, lazo_combing_2021}, for example, as used in Google's VirusTotal~\citep{shiel_2019}. 

Traditional approaches compute similarity scores between fuzzy hashes to identify shared fragments of different files. However, such similarity scores are limited in their expressiveness and, as shown by \cite{frasher}, frequently fail to detect shared fragments or even state misleading similarities between files. The fact that fuzzy hashes translate bytes into computable strings makes them especially interesting for processing on base of machine-learning techniques. With our approach, which we call \textit{\textbf{D}eep \textbf{L}earning \textbf{A}pproximate \textbf{M}atching (DLAM)}, we aim to bridge recent advances in deep learning and fuzzy hashing and demonstrate large performance improvements for the task of fragment detection by combining both concepts.

More specifically, we introduce a transformer-based~\citep{transformer} approach to detect parts of file fragments in their corresponding fuzzy hashes. We empirically show that DLAM improves fragment detection in fuzzy hashes produced by the most commonly used algorithms \ssdeep~\citep{ssdeep} and \tlsh~\citep{tlsh}, whereas traditional similarity measures fail to detect the fragments. In contrast to state-of-the-art fuzzy hashing algorithms and previous deep learning approaches, DLAM consistently classifies the fragments with an accuracy of over 90\% on multiple tested file types, namely JavaScript, PDF and XLSX.

We point out to restrict our evaluation on approximate matching schemes, which are actually used and provide an executable implementation.  In summary, we make the following contributions:
\begin{itemize}
    \item We introduce Deep Learning Approximate Matching, or DLAM in short, that combines efficient fuzzy hashing algorithms with transformer-based models from natural language processing.
    \item We show that DLAM is superior to the most common conventional approximate matching schemes \ssdeep\ and \tlsh\ and even works for small fragment sizes where traditional similarity measurements fail completely.
    \item We further demonstrate that DLAM achieves comparable results to state-of-the-art multi-resolution fuzzy hashing algorithms, such as \mrshtwo\ and \mrshcf in case of fragment detection.
    \item We provide the complete source code of DLAM on GitHub: \url{https://github.com/warlmare/DLAM}.
\end{itemize}

%% file: sections_arxiv/2_background.tex
\section{Background and Related Work}\label{sec:background}
In the following, we introduce the basics of fuzzy hashing, including \tlsh\ and \ssdeep\ as conventional fuzzy hashing algorithms on which our DLAM approach is based. We further describe the characteristics of transformer networks, which build the foundation of our machine-learning approach.

\subsection{Fuzzy Hashing and Approximate Matching}\label{sec:fuzzy_hashing}
Fuzzy hashing was first introduced in the mid-2000s in order to perform altered document and partial file matching by \citet{ssdeep}. Generally, fuzzy hashing algorithms $H$ are functions that compute a dense hash representation $H(x)$ of a file $x$. The goal is to provide similar hashes, which are also called digests, for similar inputs: ${H(x) \approx H(y) \iff	x \approx y}$. This property distinguishes them from traditional cryptographic hashes for which small input changes lead to strong output changes, also known as cryptographic diffusion or avalanche effect. Then a similarity function $sim(H(x), H(y))$ is applied, which computes the similarity between two files $x$ and $y$ based solely on their fuzzy hashes. This procedure of assessing the similarity of artifacts or patterns is known as approximate matching. Approximate matching algorithms all work in a three-step manner: they first select features from a digital artifact and generate a hash digest from these features, which then can be compared to other hashes in a third step. As for the comparison step, there are three ways in which current algorithms achieve their goal: either based on plain byte, syntactic or semantic commonalities. The matching mechanism is chosen based on the digital artifact that must be processed. As an example, malware developers often obfuscate their payload deliberately to prevent detection. Benign files might have been further compressed or rearranged through software, which makes it hard to distinguish between obfuscated malware and benign files. 

\begin{figure*}[ht]
     \centering
     \begin{subfigure}[b]{0.42\textwidth}
         \centering
         \includegraphics[height=7cm]{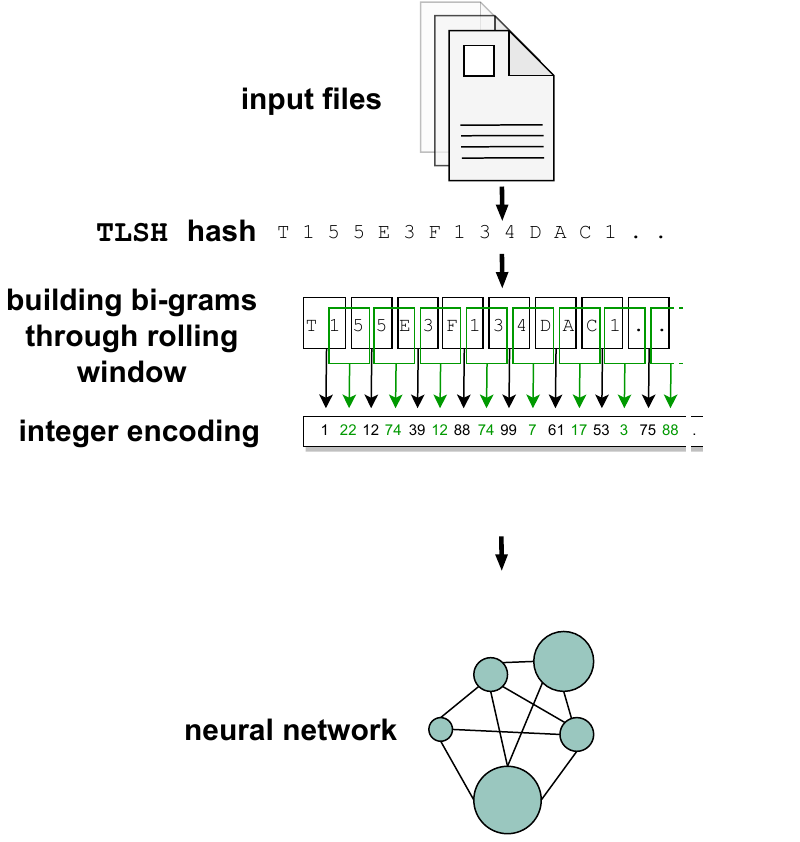}
         \caption{Feature extraction pipeline for \tlsh.}
         \label{fig:feature_extraction_tlsh}
     \end{subfigure}
     \hfill
     \begin{subfigure}[b]{0.42\textwidth}
         \centering
         \includegraphics[height=7cm]{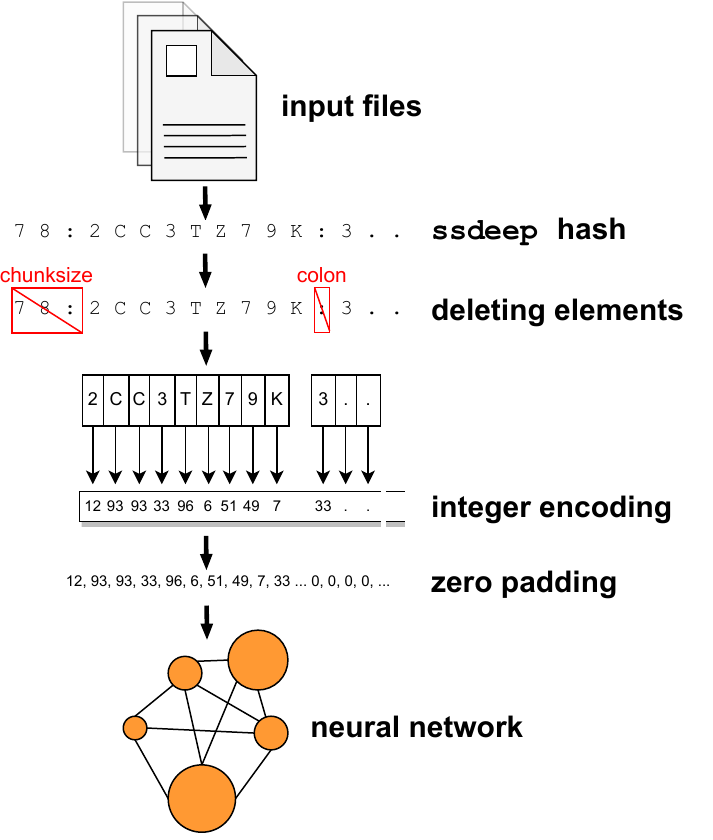}
          \caption{Feature extraction pipeline for \ssdeep.}
         \label{fig:feature_extraction_ssdeep}
     \end{subfigure}
        \caption{Feature extraction and prediction pipeline. On the left, the process for \tlsh\ hashes is depicted as it is used for both models (feed-forward and transformer). On the right, the process for \ssdeep\ hashes is depicted as it is used for creating embeddings for transformer and feed-forward models.}
        \label{fig:feature_extraction}
\end{figure*}

To deal with these issues, approximate matching research has come up with two approaches: The first approach is to create new algorithms that are more resilient and attest the similarity on a more granular level~\citep{azarafrooz_2018}. The other approach is to pre-process the artifacts before hashing them and focusing on selected parts that hold vital information~\citep{shiel_2019}. With our work, we propose a third option and detect shared commonalities in fuzzy hashes using machine learning approaches. We note that Microsoft pursues a similar direction~\citep{lazo_combing_2021}. However, this direction has so far only barely been formalized in academic research. \cref{fig:DLAM} illustrates the differences in the approaches of fragment detection of conventional approximate matching in \cref{fig:anomaly_detection_conventional} and our proposed DLAM in  \cref{fig:anomaly_detection_dlam}.

Overall, the combination of fuzzy hashes and machine learning is still in its early stages. \cite{reff2018} first proposed to train neural networks for malware detection from raw byte sequences. However, training networks with sufficient capacity on entire files is time- and resource-consuming. \cite{peiser2020} used simple feed-forward neural networks to detect malware in JavaScript files through static analysis based on features computed by fuzzy hashing algorithms and compared their results with conventional JavaScript malware classifiers, none of which rely on fuzzy hashes. They showed that simple models, trained on a small dataset with only 5,000 samples, already achieved a high detection accuracy. However, \citet{peiser2020} limited their investigation to JavaScript malware detection and only trained their models on JavaScript files that either contained malware or did not. 

In our work, we take a more holistic view and contextualize the fusion of deep learning and digital forensics based on fuzzy hashing. We further evaluate our approach on a set of different file types by injecting synthetic fragments into real-world files. We use real-world files and synthetic fragments because they allow us to study the influence that the total fragment content in a file has on the classification accuracy. Malware samples rarely come with the information how much malware is contained in a sample. 

Throughout this paper, DLAM is used in combination with the fuzzy hashes \tlsh\ and \ssdeep, which we now introduce in more detail. However, we emphasize that DLAM is, in principle, also compatible with other fuzzy hashing algorithms.

\cite{raff_2018} explained that \texttt{LZJD} could be suitable particularly for the application of machine learning. They highlight the metric space and kernel properties of the Jaccard distance that \texttt{LZJD} inherits.

\textbf{TLSH}  (Trend Micro Locality Sensitivity Hash) is a hashing algorithm first presented by \citet{tlsh}. It is specifically intended for malware detection and clustering. \tlsh\ scans the byte code of a file with a sliding window and combines 5 bytes at a time into a unit using the Pearson hash function~\citep{pearson90}, a fast non-cryptographic hashing procedure. These units are then mapped into an array of so-called buckets. Next, the digest body, a hexadecimal string, is constructed from the array of buckets by dividing it into 3 distinct quartiles q1, q2 and q3. The first three bytes of the hash are the digest header, which primarily consists of checksums, and forms together with the body a 70-character hexadecimal string. The final \tlsh\ hash is a fixed-length string with 72 characters. To compare two hashes, \tlsh\ calculates an approximation to the hamming distance between two digest bodies. The relation of two artifacts is expressed through a distance score between 0 (very similar) and 1000 (not similar).

\textbf{ssdeep} was introduced by \citet{ssdeep}. The algorithm breaks a file up into pieces using a rolling hash function, which hashes an input in a sliding window approach. Then, a non-cryptographic hash function is used to hash every piece that has been created by the rolling hash function. These smaller hashes are concatenated and form the hash signature for the whole input file. To compare two hashes, a distance measure is used that determines the correlation between the two hashes. The similarity is indicated by a score between 0 (not similar) and 100 (very similar). As the produced hashes can vary in size, the final hash contains information about the total size at the beginning of the hash. Even though the overall hash is variable in its length, the size is bounded to 148 encoded characters.

\subsection{Language Processing With Transformers}\label{sec:transformer}
To perform similarity analyses on fuzzy hashes, we interpret the binary strings as a language representation and process these strings using transformer networks. Since their introduction in 2017 by \cite{transformer}, transformers are the dominating architecture for natural language processing and are even increasingly used for image processing~\citep{dosovitskiy2021an}. Unlike previous state-of-the-art models for sequence modeling, transformers rely on the attention mechanism and remove the recurrent components of previous approaches, such as LSTMs~\citep{lstm} and GRUs~\citep{gru}. More specifically, they rely on the multi-head self-attention mechanism, which processes different representations at different positions by running an attention mechanism several times in parallel. 

The multi-head attention mechanism consists of three learned matrices: queries $Q$, keys $K$, and values $V$. The attention $A$ is computed by the dot-product $A=\frac{QK^T}{\sqrt{d_k}}$, where $d_k$ is a scaling factor based on the dimension of keys. The attention output is then computed by the weighted sum $Att(Q,K,V)=\textit{softmax}(A)\cdot V$, with the attention weights computed by the softmax function and $V$ representing the input sequence. In other words, the softmax of the attention $\textit{softmax}(A)$ is weighing the importance of the tokens in the input sequence to the task at hand. Since detecting relations between different positions in the hashes is essential for fragment detection, transformers are a natural architecture choice for this task. 

Transformers usually need large amounts of data to be trained on. Therefore, pre-training plays a crucial role in their performance. \cite{bert} proposed bidirectional encoder representations from transformers (BERT), a masked language model, which introduced a novel pre-training procedure. To train the masked language model, randomly selected words in the input sequences of the training corpus are masked out. The model then has to predict the masked word based on its surrounding context. 
A pre-trained model can then be fine-tuned for other NLP tasks with only a single additional output layer added. 

After demonstrating impressive results in natural language tasks, the transformer architecture has been transferred to other domains as well. These include computer vision with vision transformers~\citep{vit}, speech recognition with convolutional-augmented transformers, so-called conformers~\citep{conformer}, and reinforcement learning with decision transformers~\citep{decision_transformer}. With our work, we take the application range of transformers one step further and combine them with fuzzy hashing algorithms to improve approximate matching.

%% file: sections_arxiv/3_experiments.tex
\section{Transformer-Boosted Approximate \\Matching}\label{sec:transformer_boosted}

\begin{figure*}[ht]
    \centering
    \includegraphics[width=0.95\textwidth]{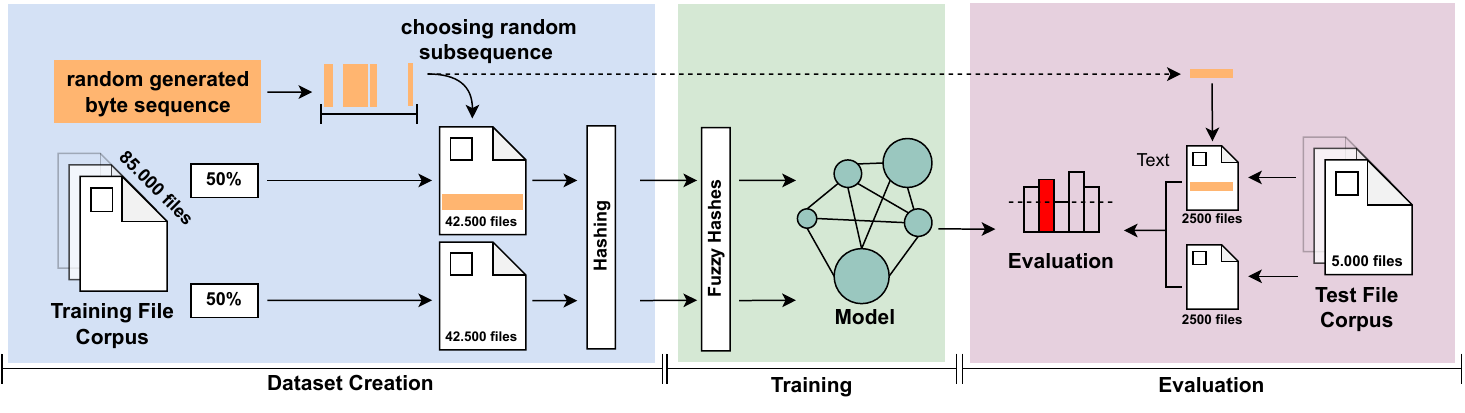}
    \caption{The three stages of our training and evaluation pipeline.}
    \label{fig:pipeline}
\end{figure*}

In machine learning assisted common fragment detection, the model should learn what constitutes a regular file and what constitutes a file containing a particular element we want to detect. In this work, we introduce DLAM to improve the detection behavior of traditional fuzzy hashing algorithms. Our approach consists of two steps. For each input file $x$ to check, we first compute a dense hash representation $H(x)$ using a fuzzy hash algorithm $H$. The fuzzy hashes serve in this case as an intermediate representation of the files and can be seen as an input pre-processing step. The hashes reduce the complexity and length of the original files and make them processable by deep learning models. 

We investigate supervised fragment detection, which comes down to a binary classification of files to predict if some fragment is present in an input file. The hashes and binary labels, which indicate whether a fragment was hidden in the input file or not, are then provided to a deep learning model, in our case a transformer, for training. Since the hashes are in a byte string representation, they are tokenized and interpreted as a sequence to train the model. Because the order of the bytes is crucial to the classification task, we added a simple absolute positional encoding to the sequence to indicate the sequence order to the transformer model after tokenizing the hash strings.

As stated in \cref{sec:fuzzy_hashing}, \texttt{ssdeep} produces hashes of variable length. As \cref{fig:feature_extraction_ssdeep} illustrates, all hashes produced by \texttt{ssdeep} are padded with zeros to unify the length of the hashes, allow parallel training of the models, and apply attention masking to train the transformer model. Attention masking is used to prevent the model from attending to the padding tokens and only paying attention to the informative tokens of the hash values. As different training samples have different sizes, masking out the padding tokens prevents the model from overfitting the training data and predicting the appearance of the fragment based on the number of padding tokens.

\subsection{Experimental Protocol}
In the following, we will describe the experimental setting we used to evaluate our approach. The pipeline for our experiments is visualized in \cref{fig:pipeline}.\\

\textbf{Fuzzy Hashing Algorithms:} In our experiments, we focus on \ssdeep\ and \tlsh, as both are widely used and form the de-facto standard for malware detection. As such, they are both part of Google's VirusTotal, Maltego and other actively used virus detection platforms. Thanks to the fact that transformers can process inputs of arbitrary length, our approach is also applicable to fuzzy hashes of arbitrarily length, such as those generated by \sdhash\ or \mrshtwo\ (\mrshcf\ is currently unable to provide its fuzzy hashes as readable strings). However, multi-resolution hashes are many times longer than \tlsh\ and \ssdeep, which are limited in their maximum length. This potentially unlimited length is a drawback when processing fuzzy hashes by machine learning. In their paper, \cite{peiser2020} also apply a simple feed forward network to the \sdhash\ algorithm, which is also theoretically unbounded in length. However, they use the method of count vectorization, which means that important information such as the relative position of the segments, which we consider crucial, is lost and therefore abandon the approach adopted by \cite{peiser2020}.

The feature extraction pipeline for both fuzzy hashes \ssdeep\ and \tlsh\ is visualized in \cref{fig:feature_extraction}. According to \cite{peiser2020}, when using a rolling-window approach for tokenizing the fuzzy hashes produced by \tlsh, the classification accuracy is increased with a feed-forward network. In our experiments, the rolling-window approach did not influence the classification accuracy with a transformer. However, to ensure comparability to \cite{peiser2020}, we still used a rolling-window approach for tokenizing the fuzzy hashes produced by \tlsh.

Contrary to the findings of \cite{peiser2020}, who used a similar rolling window approach on \ssdeep\ for training their feed-forward network as well, we could not confirm that such an approach improved the classification performance of the feed-forward models when working with \ssdeep. Therefore, we applied a simple per-char integer encoding on the \ssdeep\ hash, as can be seen in \cref{fig:feature_extraction_ssdeep}. We also removed the chunk size and colons from the \ssdeep\ hashes, for both of the model architectures. During our experiments, we noticed that the retention of these elements leads to a worse classification accuracy by the models. \\

\textbf{Testbed:}
The test environment in which all our tests were conducted is as follows. We use a virtual machine equipped with 16x Intel Xeon Platinum 8280 CPU @ 2.70GHz, 2x 16 GB DDR4-3200 RDIMM ECC, 800 GiB SSD, Ubuntu 21.10 Linux 5.13.0-40-generic x86\_64. In addition, three Nvidia Tesla V100s with 32 GB VRAM are used for the training of the deep learning assisted approximate matching models. All modules for the creation of the training data as well as the machine learning models and the evaluation were realized in Python 3.8. Although this setup is quite extensive, it is still not sufficient when dealing with fuzzy hashes of variable length. We suggest this as future research with more powerful hardware. \\

\textbf{Data Selection:}
To investigate the extent to which the classification performance of the models depends on the type of input file, we selected common file types, namely JavaScript, PDF and XLSX, for training and evaluation. \cite{goebel_2021} showed that the accuracy of approximate matching indeed varies depending on the file type. Therefore, we trained and evaluated one model per file type. Since no single file corpus exists that is big enough to allow for training and evaluation of models on all file types, multiple corpora were used. We combined the \textit{Govdocs corpus}~\citep{govdocs}, \textit{SRILabs JavaScript corpus}~\citep{raychev_2016} and the \textit{FUSE corpus}~\citep{barik_2015} and used 100,000 samples for each of the file types. This number of samples per file type represents the maximum amount of files that could be assembled with a reliable provenience, meaning that they originate from corpora that have been screened for duplicates and are balanced in size. 

For evaluation of the models, 5,000 samples for each file type from the \textit{NapierOne corpus}~\citep{davies_2022} were used. For training and evaluation, only files that were large enough to be reliably hashed by \tlsh\ were selected. Since the files for training and evaluation were collected from different sources, the risk of having duplicate files in the training and test sets is minimal.

\begin{table*}[ht]
\small
\resizebox{\textwidth}{!}{
    \begin{tabular}{l c c c c c | c c c c} 
    \toprule
    && \multicolumn{4}{c|}{\textbf{Traditional Fuzzy Hashing}} & \multicolumn{4}{c}{\textbf{DLAM (Ours)}} \\
    & \textbf{} & \textbf{mrsh-cf} & \textbf{MRSH-v2} & \textbf{ssdeep} & \textbf{TLSH} & \textbf{ssdeep (FF)} & \textbf{TLSH (FF)} & \textbf{ssdeep (TF)} & \textbf{TLSH (TF)}  \\ 
    \midrule
    \parbox[t]{2mm}{\multirow{5}{*}{\rotatebox[origin=c]{90}{\textbf{JS}}}}

    & Accuracy (\%) & 81.54    & 67.38 & 50.02 & 50.02 & \textbf{92.70}    & 82.08 & \textbf{92.70}    & \textbf{87.90}    \\
    & TPR (\%)      & 0.63     & 0.83  & 0.00  & 0.00  & 0.86              & 0.75  & 0.86              & 0.84              \\
    & FPR (\%)      & 0.00     & 0.05  & 0.00  & 0.00  & 0.01              & 0.11  & 0.01              & 0.08              \\
    & TNR (\%)      & 1.00     & 0.95  & 1.00  & 1.00  & 0.99              & 0.89  & 0.99              & 0.92              \\
    & FNR (\%)      & 0.37     & 0.17  & 1.00  & 1.00  & 0.14              & 0.25  & 0.14              & 0.16              \\
    
    \midrule
    \parbox[t]{2mm}{\multirow{5}{*}{\rotatebox[origin=c]{90}{\textbf{PDF}}}}
    & Accuracy (\%) & \textbf{97.3} & \textbf{95.44}    & 50.04 & 49.98 & 79.10 & 78.42 & \textbf{94.34}    & 82.60 \\
    & TPR (\%)      & 0.95          & 0.93              & 0.00  & 0.00  & 0.75  & 0.74  & 0.93              & 0.83  \\
    & FPR (\%)      & 0.00          & 0.02              & 0.00  & 0.00  & 0.16  & 0.17  & 0.04              & 0.17  \\
    & TNR (\%)      & 1.00          & 0.98              & 1.00  & 1.00  & 0.84  & 0.83  & 0.96              & 0.83  \\
    & FNR (\%)      & 0.05          & 0.07              & 1.00  & 1.00  & 0.25  & 0.26  & 0.07              & 0.17  \\

    \midrule
    \parbox[t]{2mm}{\multirow{5}{*}{\rotatebox[origin=c]{90}{\textbf{XLSX}}}}
    & Accuracy (\%) & \textbf{96.78}    & 88.22 & 52.54 & 51.17 & 93.80 & 90.28 & \textbf{97.36}    & \textbf{90.74}    \\
    & TPR (\%)      & 0.94              & 0.77  & 0.05  & 0.03  & 0.89  & 0.89  & 0.96              & 0.92              \\
    & FPR (\%)      & 0.00              & 0.00  & 0.00  & 0.04  & 0.01  & 0.08  & 0.01              & 0.11              \\
    & TNR (\%)      & 1.00              & 1.00  & 1.00  & 0.96  & 0.99  & 0.92  & 0.99              & 0.89              \\
    & FNR (\%)      & 0.06              & 0.23  & 0.95  & 0.97  & 0.11  & 0.11  & 0.04              & 0.08              \\

    \bottomrule
    \end{tabular}
    }
    \caption{Comparison of our results for fragment detection in JavaScript (JS), PDF, and XLSX files for traditional similarity-based fuzzy hashing algorithms (left side) and DLAM with feed-forward (FF) and transformer (TF) models (right side). Whereas traditional similarity metrics fail for the fixed-sized fuzzy hashes generated by TLSH and ssdeep, our DLAM approaches significantly increase the performance and are even more accurate than state-of-the-art fuzzy hashing algorithms with variable hash sizes for JavaScript and XLSX files. We highlighted the best 3 accuracy values for each file type.}
    \label{tab:anomaly_detection_results}
\end{table*}

\textbf{Evaluation Metrics:} We used various standard metrics for evaluating our experimental results. We measured the accuracy (Acc), the false-positive rate (FPR), the true-positive rate (TPR), the false-negative rate (FNR), and the true-negative rate (TNR). Regarding the application of our approach in digital forensics for malware or sensitive file detection, false-positive and false-negative predictions are the most relevant performance indicators. False-negative predictions represent all files that a filter failed to detect. Any false-negative prediction can mean a compromise of systems with potentially devastating consequences. False-positive predictions, in turn, are fatal misclassifications in the context of malware detection because they impair processes for no reason and lead to a sustained loss of confidence in the technology by the users. 

\begin{figure*}[ht]
    \centering
    \begin{subfigure}[b]{0.9\textwidth}
        \includegraphics[width=\linewidth]{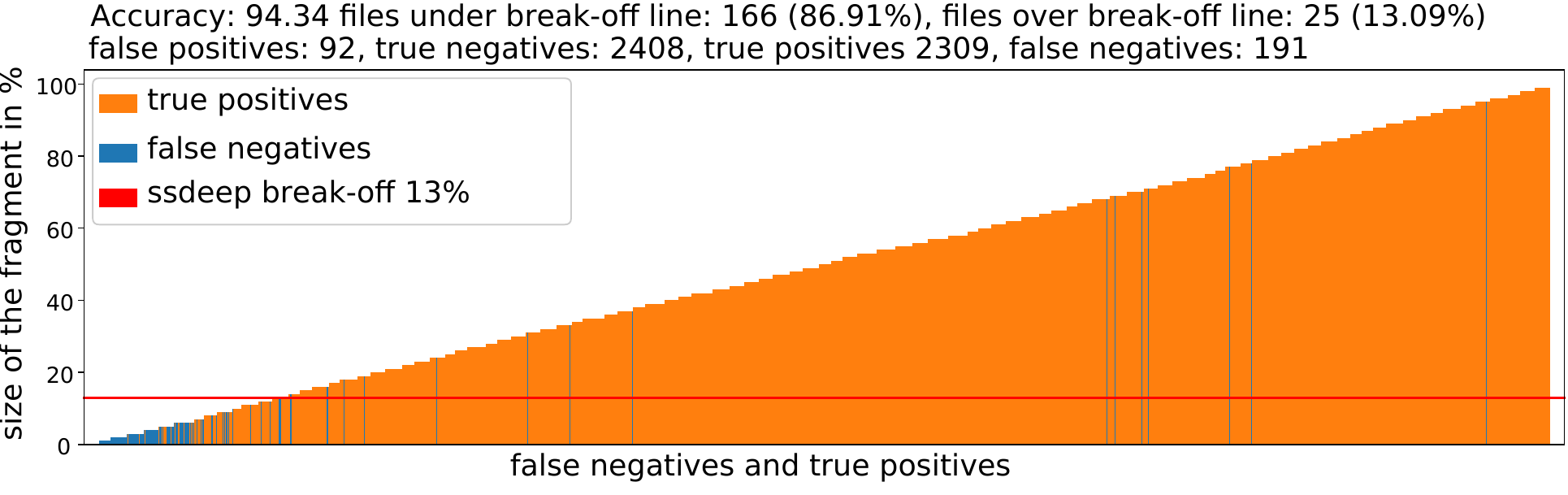}
        \caption{\ssdeep}
        \label{fig:false_negatives_true_positives}
    \end{subfigure}
    \begin{subfigure}[b]{0.9\textwidth}
        \includegraphics[width=\linewidth]{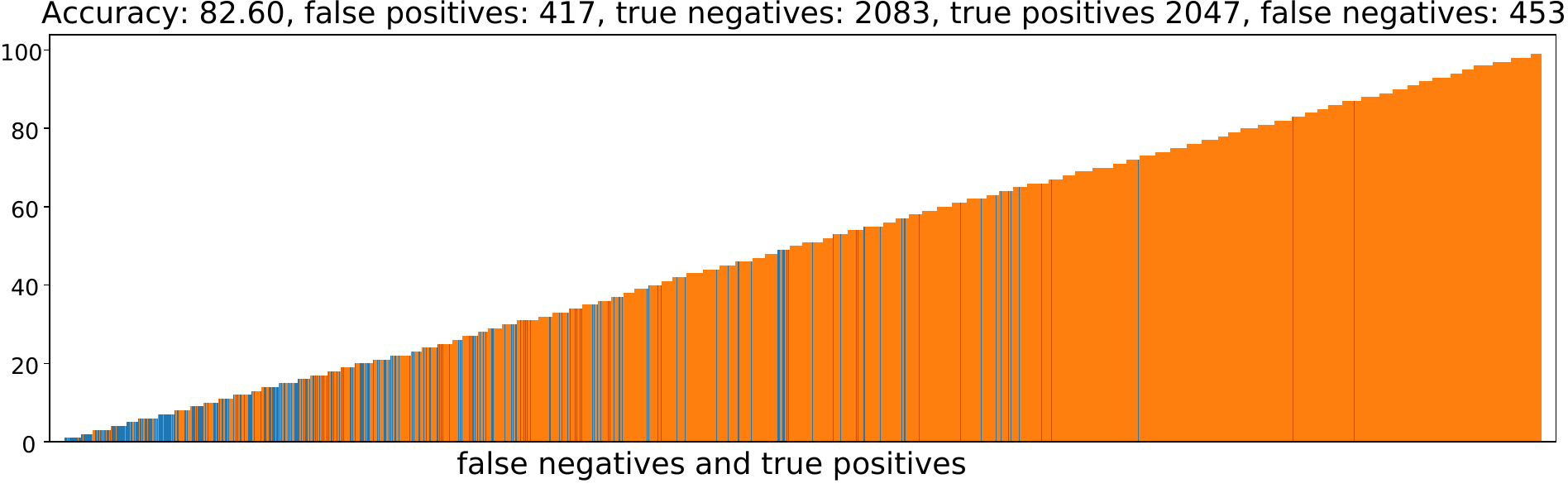}
        \caption{\tlsh}
        \label{fig:false_negatives_true_positives_tlsh_transformer}
    \end{subfigure}
    \caption{Classification results of DLAM with a transformer model for all PDF files that carry an to-be-detected fragment within them. The blue bars symbolize false negatives, while the yellow ones symbolize true positives. The height of each bar is proportional to the size of the to-be-detected fragment within the file. There are 2,500 bars in each graph. The red line in \cref{fig:false_negatives_true_positives} symbolizes the so-called \textit{ssdeep-break-off}. It corresponds to the minimum amount of relation that \ssdeep\ can conventionally approximately match.} 
\label{fig:transformation_results}
\end{figure*}

\textbf{Training Setup:}
The architecture of the transformer model relies on the \textit{TinyBERT} architecture by \cite{tinybert}. The model is a smaller, distilled version of the widely used, pre-trained language model BERT~\cite{bert}. Compared to BERT, TinyBERT models have significantly less number of parameters, which improves training and inference time noticeably. All our experiments were implemented in the deep-learning framework PyTorch~\citep{pytorch}. We note that a single GPU, as, for example, provided by Google Colab for free, is sufficient to train our models on 100,000 files in 20 minutes. Training on more powerful, data-center-grade GPUs, e.g., an NVIDIA A100 GPU, can speed up the training time significantly. Furthermore, training can also be scaled across multiple GPUs, allowing for larger batch sizes to accelerate the training process even more. Adaptive sampling strategies to create a representative subset of available data might also be used to reduce the effective number of samples in the training data.

Following the original TinyBERT architecture, our transformer encoder model has 12 attention heads and 4 hidden layers. The intermediate feed-forward layer size in the encoder is 1200. The dimensionality of the encoder and pooling layers is 312. This model has a sufficient size for our purposes, as it needs less training data than larger transformers. We further initialized our model with pre-trained weights trained on a natural language task and found that it boosts the performance even if fine-tuned on abstract hashes compared to training a randomly initialized model from scratch. The output layer is a standard feed-forward layer with two output neurons.

For comparison reasons, we also trained a simple feed-forward neural network, whose architecture was inspired by the work of \cite{peiser2020}. The model consists of three fully-connected layers, followed by batch normalization layers~\citep{batchnorm} and ReLU activations. It also contains a dropout layer with dropout probability $p=0.125$ after the first block.

We trained all networks using the \textit{Adam optimizer}~\citep{adam} with a fixed learning rate of $10^{-3}$, a batch size of 1,024 tokenized hashes for our transformer model (TinyBERT), and a smaller batch size of 512 for the feed-forward network. We further used 15\% of the training data as a validation set and performed early stopping, meaning that we stopped training after the validation loss started to increase. We then used a standard cross-entropy loss function for optimization:
\begin{equation*}
\mathcal{L}_{CE}=-\sum_{i=1}^{n} y_i \cdot \log (\hat{y}_i).
\end{equation*}
Here, $y_i\in {0,1}$ denotes the ground-truth binary indicator if the corresponding file contains a to-be-found fragment, and $\hat{y}_i\in [0, 1]$ denotes the predicted probability of $y_i=1$. We emphasize that the model could also be designed with a single output and optimized with a binary cross-entropy loss. However, by designing the model for multi-classification, we could easily extend it for additional classes, if necessary. 

For each of the three file type settings, the models were trained on 85,000 files and evaluated on 5,000 files. This is visualized in \cref{fig:pipeline}. Note that the training and evaluation files come from different datasets. Into half of the training and evaluation files, a distinct fragment in the shape of a randomly generated byte sequence was inserted. Even though the same randomly generated byte sequence was used for creating all the files containing a fragment, the specific elements that were copied were chosen at random. In addition, the inserted fragment could occupy between {1\%} and {99\%} of the original size of the file.

To investigate how traditional fuzzy hashes perform the approximate matching task without the help of machine learning, the common fuzzy hashing algorithms \ssdeep, \tlsh, \mrshcf, and \mrshtwo\ were used to compare each evaluation file with the entire randomly generated to-be-found fragment sequence. If their resulting similarity scores were greater than zero, they were considered a \textit{positive} prediction. If the similarity score was zero, this was considered a \textit{negative} prediction. \\

\subsection{Experimental Results}
We now want to empirically investigate the following research questions: \\
\begin{itemize}
    \item[\textbf{Q1:}]  Can DLAM detect file fragments more precisely than traditional approximate matching? 
    \item[\textbf{Q2:}]  Does the type of file containing the fragment affect detection with DLAM?
    \item[\textbf{Q3:}]  Does DLAM with transformer networks perform better than with simple neural networks?
    \item[\textbf{Q4}:]  Can DLAM compensate for weaknesses in conventional approximate matching with fuzzy hashing?
\end{itemize}

\begin{table*}[ht]
    \centering
    \small
    \resizebox{0.9\linewidth}{!}{%
    \begin{tabular}{c c c c c |c c} 
    \toprule
    & \multicolumn{4}{c}{\textbf{Traditional Fuzzy Hashing}} & \multicolumn{2}{c}{\textbf{DLAM (Ours)}} \\
    \textbf{Multiplication factor} & \textbf{mrsh-cf} & \textbf{MRSH-v2} & \textbf{ssdeep} &  \textbf{TLSH} & \textbf{ssdeep (TF)} &  \textbf{TLSH (TF)} \\ 
    \midrule
        x 1  & 97.54 & 76.13 & 50.0 & 50.0 & 91.09 & 94.94 \\
        x 2  & 92.19 & 74.37 & 50.0 & 50.0 & 81.68 & 93.57 \\
        x 4  & 91.99 & 72.67 & 50.0 & 50.0 & 63.56 & 93.25 \\
        x 8  & 91.69 & 76.47 & 50.0 & 50.0 & 53.55 & 93.07 \\
        x 16 & 91.49 & 73.73 & 50.0 & 50.0 & 53.55 & 93.95 \\
        x 32 & 88.21 & 71.11 & 50.0 & 50.0 & 50.0  & 93.59 \\
    \bottomrule
    \end{tabular}
    }
    \caption{Prediction accuracy for fragment detection with traditional fuzzy hashing algorithms with variable (\mrshcf\ and \mrshtwo) and fixed (\ssdeep\ and \tlsh) hash sizes, and our DLAM approach combining \ssdeep\ and \tlsh\ with a transformer network. Per row, 5,000 files with a file size of 5,000 bytes were created. Half of them contained between {1\%} and {99\%} of fragment bytes within them. All files were concatenated according to the respective multiplication factor and then had to be classified. The results are averaged over 10 test runs.}
    \label{tab:digest_comparison_impediment_model_results}
\end{table*}

\bigskip
\textbf{Q1: DLAM meets accuracy of state-of-the-art fuzzy hashing algorithms.} 
We state our experimental results, which compare our DLAM approach to common and state-of-the-art fuzzy hashing algorithms in \cref{tab:anomaly_detection_results}. The table includes the accuracy, false-positive rate, true-positive rate, false-negative rate, and true-negative rate in percentage. While \ssdeep\ and \tlsh\ on their own completely fail in detecting the files with injected fragments for all file types, the state-of-the-art hashing algorithms \mrshcf\ and \mrshtwo\ are still able to reliably identify the fragments in most cases. This is especially interesting since the files to be classified have the same length as the fragment, which in theory is an easy setting for \ssdeep\ and \tlsh\ as was shown by \citet{martinez2020improved}. Unlike \ssdeep\ and \tlsh, \mrshcf\ and \mrshtwo\ are multi-resolution hashes, which can map many more features into the hashes. As a result, the length of the hashes is not limited to a maximum size, which increases accuracy for similarity matching since less information about the files is lost.

As for the deep learning approaches, our approach is on par with the feed-forward model of \citet{peiser2020} for JavaScript files, but significantly outperforms the feed-forward model on PDF and XLSX files by 15.24\% and 3.56\%, respectively. Our transformer-based approach generally performs better when using the hashes of \ssdeep\ than with using the \tlsh\ hashes. The superiority of \ssdeep\ over \tlsh\ for approximate matching with real-world file types was also pointed out by \citet{frasher}.

Our approach also outperforms the state-of-the-art multi-resolution fuzzy hashing functions \mrshcf\ and \mrshtwo\ on the JavaScript files. While the accuracy of the approach of \citet{peiser2020} is 18.2\% lower than the state-of-the-art hashes, the accuracy of our approach is only 2.96\% below the \mrshcf\ hash on PDF files. On XLSX files, the transformer-based DLAM approach even outperforms the \mrshcf\ hash and achieves an accuracy of 97.36\%. This is especially interesting since our transformer-based approach enables fragment detection based on fuzzy hashes like \ssdeep\ and \tlsh, which scale much better than the mentioned multi-resolution hashes with only moderate accuracy losses when using DLAM. As mentioned at the beginning, \ssdeep\ and \tlsh\ are used for malware detection in practice. Their short length means that they map large amounts of data to a fraction of the disk space that \mrshtwo-hashes would take up for the same data.

\cref{fig:false_negatives_true_positives} and \cref{fig:false_negatives_true_positives_tlsh_transformer} show the false negatives (as blue bars) and the true positives (yellow bars). The height of the bars represents the proportion of the fragment in the total file. In both figures, the accumulation of blue bars in the left half of the graphs shows that the smaller the proportion of to-be-found fragment in a file, the higher the risk of a false-negative classification by the model.    

With \ssdeep, the model detects the to-be-found fragments in the hashes much more consistently, even for small fragment sizes. The red line in \cref{fig:false_negatives_true_positives} is an approximation of the minimum amount of similarity that two files need to have in common, in order for \ssdeep\ to approximately match them. \cite{frasher} discovered that between 9\% and 13\% was the minimum amount of common content that two randomly generated files of size 2048~KB needed to possess in order for \ssdeep\ to find any relation. Even though this minimum was never confirmed for PDF files specifically, our experimental results support this theory.  

The fact that almost 87\% of all false negatives have a smaller fragment than 13\% means that the smaller the fragment, the less likely it is to be classified correctly. The false negatives in \cref{fig:false_negatives_true_positives_tlsh_transformer} are more widespread and cannot primarily be attributed to a small fragment size, as for \ssdeep\ in \cref{fig:false_negatives_true_positives}.  
\cite{frasher} stated that \tlsh\ was less precise than \ssdeep\ when tested on various file types in their \textit{needle in a haystack} test case. An observation that we could confirm with our tests. The classification accuracy of any model was consistently lower for \tlsh\ than that of \ssdeep. Our experiments show that working consciously with DLAM and \ssdeep\ within these boundaries -- only classifying files with to-be-found content of more than 13\% -- even leads to a higher classification accuracy than with \mrshcf\ or \mrshtwo.
\\

\textbf{Q2: DLAM reliably detects fragments for all file types.} Our results in \cref{tab:anomaly_detection_results} further illustrate that the accuracy with which DLAM detects fragments in files seems to be more stable in face of different file types. Using a feed-forward network leads to highly fluctuating accuracy values for each file type. The same can be seen for the multi-resolution hashes. Even though their accuracy is very high on PDF and XLSX files, there is a high loss of accuracy on JavaScript files, indicating that the file types have a high impact on the performance of state-of-the-art fuzzy hashes. Our transformer-based DLAM approach, on the other hand, is much more stable and the accuracy for the fragment detection is consistently above 90\% for all three file types when using \ssdeep\ hashes. It, therefore, offers higher reliability in its predictions for different domains than fuzzy hashing algorithms on their own. \\

\textbf{Q3: Transformers outperform simple neural networks for fragment detection.}
As can be seen in \cref{tab:anomaly_detection_results}, our DLAM approach with \ssdeep\ consistently outperforms the neural networks with \tlsh\ hashes. Using \tlsh, the model produces more false-negatives than with \ssdeep, as can be seen in \cref{fig:false_negatives_true_positives_tlsh_transformer} and \cref{fig:false_negatives_true_positives}.
The results in \cref{tab:anomaly_detection_results} also show that a combination of transformers and \ssdeep\ consistently outperforms all approaches that are based on simple feed-forward networks. So choosing a more complex architecture with a much higher number of parameters pays off in this case. 

However, even if feed-forward networks perform worse than transformer models in most cases, they still achieve a decent performance. Consequently, for applications with limited computational power available or to speed up inference time, feed-forward networks might still be a reasonable architecture to choose.
\\

\textbf{Q4: DLAM increases the robustness when working with fuzzy hashes.} 
One way to degrade the performance of traditional fuzzy hashes is to concatenate the same file multiple times before feeding them into the algorithms. We show that our DLAM approach strongly increases the detection robustness of fuzzy hashes. We concatenated the same file a pre-defined number of times and then calculated the hashes for the composed file.
\cref{tab:digest_comparison_impediment_model_results} states our results and illustrates that DLAM can indeed compensate for the known low resilience against repetitive content. 

As also shown in previous research \citep{frasher}, the similarity scores of \ssdeep\ and \tlsh\ could be lowered to 0 when comparing a single instance of a file with another file that consisted of multiple concatenations of the first one. In the case of \tlsh, a 16-fold concatenation of the same 5,000 bytes was sufficient to lower the similarity score to 0, leading to a prediction accuracy of 50\%, which comes down to random guessing in a binary prediction problem. However, our results further show that \tlsh, in combination with a transformer network, is still able to detect files with injected fragments and benign files with higher accuracy than \mrshcf\ and \mrshtwo\ in face of a repetition. These results also indicate that there is much more information present in \tlsh\ hashes than can be processed by simple similarity measurements.

%% file: sections_arxiv/4_discussion.tex
\section{Discussion and Limitations}\label{sec:discussion}

With this work, we demonstrated the large potential that lies in the combination of traditional fuzzy hashing algorithms and recent advances in deep learning. For the task of supervised fragment detection, our DLAM approach achieves accuracy values close to those of multi-resolution fuzzy hashes. Furthermore, the fuzzy hashes \ssdeep\ and \tlsh\ used in combination with DLAM are far more scalable due to more effective compression procedures. For example, an unbounded \mrshtwo\ hash compresses 68x worse than a \ssdeep\ hash bounded to 148 bytes in length. With our DLAM, a comparatively small hash length can then still be classified with high accuracy, comparable to conventional approximate matching with \mrshcf\ and \mrshtwo. For the supervised fragment detection in JavaScript, DLAM achieves even higher accuracy values than all state-of-the-art \mrshcf\ and \mrshtwo. Furthermore, DLAM also increases the resilience against file repetition, where \ssdeep\ and \tlsh\ alone fail to provide meaningful similarity scores.
\newpage
Based on our empirical results, we recommend combining DLAM with \ssdeep\ over \tlsh, since \ssdeep's weaknesses can be narrowed down much more precisely. Similarities smaller than 15\% are no longer safely detectable and should therefore be avoided when classifying with DLAM and \ssdeep. \tlsh, on the other hand, is generally less precise, and possible false predictions are much more difficult to predict. A way to overcome the limitations of \ssdeep\ with regard to smaller similarities could lie in the work of \cite{shiel_2019} on improving file-level fuzzy hashes for malware variant classification. The so-called section-level hashing means that digital artifacts are not hashed and compared in their entirety, but only in sections. By compartmentalization – dividing the digital artifacts into equally sized blocks and hashing each one – a much more granular search for possible similarities can be performed. So one input file is no longer represented by a single hash but by multiple ones, each representing the exact same amounts of bytes. 

We limited our analyses to algorithms that produce hashes with a fixed-sized (\tlsh) or a maximum length (\ssdeep). Other algorithms, such as \mrshtwo, produce large hashes with potentially infinite lengths. Since multi-resolution hashes can theoretically have any length, transformers that can process inputs of any length are applicable in theory. \cite{peiser2020} were able to process, \texttt{sdhash} which produces a limitless hash with a so-called bag of words approach. However, this has the disadvantage that all positional information on where sequences are located in a hash is lost. Which is why it was not adapted for \mrshtwo\ in this work. \\

\subsection{Practical Application}

The main practical applicability of this work is illustrated here once again with an example. DLAM can be understood as a building block to gain significantly more visibility in the context of incident response or endpoint detection and response. As \cite{lazo_combing_2021} describes, files can be hashed on endpoints and these hashes are matched by the EDR solution with known hashes, i.e., malware IOCs. DLAM now enables the targeted training of machine learning networks to detect specific malware signifiers in fuzzy hashes (i.e., ransomware, adware, etc.). Until now, this required either a large number of cryptographic hashes, which must cover all polymorphic forms of a malware, or fuzzy hashes, which can only express a shared content with known malware files by the similarity score. However, this is rather unreliable as the shared content might just be coincidental noise. With DLAM, machine learning models can be trained to classify files based only on specific malware sequences in the fuzzy hashes. The models are trained to distinguish between benign files and malware. Our results show that this is already possible with the common fuzzy hashes \ssdeep\ and \tlsh. However, this is all the more impressive as these two hashing algorithms produce very short hash strings on which DLAM can be trained. Applied onto \ssdeep\ and \tlsh, DLAM can achieve classification accuracies that match and even surpass those of fuzzy hashing algorithms that are unlimited in length (like \mrshcf\ and \mrshtwo). As \cite{lazo_combing_2021} has shown, an approach like DLAM can be used to discover a previously unknown strand of malware altogether.

%% file: sections_arxiv/5_conclusion.tex
\section{Conclusion and Future Work}\label{sec:conclusion}

The goal of our work is to open up the research on machine learning and fuzzy hashes to the broader scientific community. With deep learning approximate matching, or DLAM in short, it is now possible to contextualize a promising fusion of the two research areas of deep learning and digital forensics based on fuzzy hashing. Our application of transformer models in this context leads to higher classification accuracy than other current approaches. More precisely, we empirically demonstrated that applying DLAM to efficient standard fuzzy hashes, such as \ssdeep\ and \tlsh, creates a powerful application that rivals the best-performing fuzzy hashes \mrshtwo\ and \mrshcf\ in terms of supervised file fragment detection. By adjusting the size of the inserted fragment and the file type, we were able to show the impact of these two factors, both of which are important when using DLAM in practical applications for data loss prevention or malware detection. The full potential that DLAM offers for digital forensics and cybersecurity becomes even more apparent, recalling that, \ssdeep\ and \tlsh\ are very compact and many times smaller than multi-resolution hashes. This paper proves that machine learning is a powerful enabler that enhances the practical applications of fuzzy hashes well beyond the approximate matching.

As was explained previously, we could not apply DLAM (Transformer) on fuzzy hashes with potentially unlimited length due to hardware constraints. This does not mean that future research should shy away from exploring this area, we leave the realization open for future research. 

Another interesting avenue is to extend DLAM beyond the binary classification of anomalies, which we explored in this work. We expect DLAM to also improve results for unsupervised fragment and outlier detection, content retrieval, and multi-class file classification based on fuzzy hashes, but leave the empirical proof for future work.

\newpage
\paragraph{Acknowledgements} This work was supported by the German Ministry of Education and Research (BMBF) within the framework program ``Research for Civil Security'' of the German Federal Government, project KISTRA (reference no. 13N15343). It also benefited from the National Research Center for Applied Cybersecurity ATHENE, a joint effort of BMBF and the Hessian Ministry of Higher Education, Research, Science and the Arts (HMWK). 